\documentclass[camera,copyright,creativecommons]{eptcs}

\usepackage[british]{babel}
\usepackage{alltt}
\usepackage{amsmath}
\usepackage{amssymb}
\usepackage{bussproofs}
\usepackage{graphicx}
\usepackage{hyperref}
\usepackage{stmaryrd}
\usepackage{xspace}

\newtheorem{definition}{Definition}[section]

\newtheorem{theorem}[definition]{Theorem}
\newtheorem{proposition}[definition]{Proposition}
\newtheorem{lemma}[definition]{Lemma}

\newtheorem{theor*}{Theorem}
\newtheorem{lem*}[theor*]{Lemma}
\newtheorem{prop*}[theor*]{Proposition}
\newtheorem{corol*}[theor*]{Corollary}
\newtheorem{examp*}{Example}

\mathcode`\.="213A 
\mathcode`\|="226A 

\newcommand*{\ie}{\emph{i.e.}\xspace}
\newcommand*{\ea}{\emph{et al.}\xspace}

\newcommand*{\ok}{\ensuremath{\opnsf{ok}}}

\newsavebox{\figurebox}
\newenvironment{myfigure}
{\begin{figure}[tp]\begin{lrbox}\figurebox\begin{minipage}{0.96\textwidth}}
{\par\unskip\vspace{-1em}\vspace{1pt}\null\end{minipage}\end{lrbox}\framebox[\textwidth][c]{\usebox\figurebox}\end{figure}}
\let\myparagraph\paragraph

\newcommand*{\bnf}{::=\:}
\newcommand{\alt}{\:|\:}
\newcommand*{\pars}{\;|\;}
\newcommand{\eqdef}{\ensuremath{\stackrel{\operatorname{def}}{\Relbar}}}
\newcommand{\bdef}{\mathbin{\eqdef}}

\newcommand*{\set}[1]{\{#1\}}
\newcommand*\0\emptyset

\newcommand*\rulename[1]{\LeftLabel{\scshape (#1)}}
\newcommand*\rrulename[1]{\RightLabel{\scshape (#1)}}

\newcommand*\subsess\leqslant
\newcommand*\supsess\geqslant
\newcommand*\subgen\leq
\newcommand*\supgen\geq
\newcommand*\eqgen\simeq
\newcommand*\subst[3]{#1\left[#2/#3\right]}
\newcommand*\structpre\preccurlyeq
\newcommand*\struct\cong

\newcommand*{\opnsf}[1]{\ensuremath{\mathop{\mathsf{#1}}}}
\newcommand*{\tlbs}[1]{\opnsf{tlbs}({#1})\xspace}
\newcommand*{\var}{\mathcal{V}}
\newcommand*{\dom}[1]{\opnsf{dom}({#1})\xspace}
\newcommand*{\fn}[1]{\opnsf{fn}({#1})\xspace}
\newcommand*{\fv}[1]{\opnsf{fv}({#1})\xspace}
\newcommand*{\wf}[1]{\opnsf{WF}({#1})\xspace}
\newcommand*{\wfg}{\wf\Gamma}
\newcommand*{\lin}[1]{\opnsf{LIN}({#1})\xspace}
\newcommand*{\linop}{\opnsf{LIN}}
\newcommand*{\ling}{\lin\Gamma}
\newcommand*{\okg}{\ok(\Gamma)}

\newcommand*\taumove{\mathbf{t}}
\newcommand*\wideexclam{\makebox[.25em]{!}}
\newcommand*\widequestion{\makebox[.25em]{?}}
\newcommand*\tsl[1]{{\ensuremath\left\llbracket#1\right\rrbracket}}
\newcommand*\tsls[1]{{\ensuremath\llparenthesis#1\rrparenthesis}}
\newcommand*\sel\triangleleft
\let\phi\varphi\let\epsilon\varepsilon\let\emptyset\varnothing
\newcommand*\branch[1]{\triangleright\left\{#1\right\}}
\newcommand*\echoice[1]{\mathop{\&}\left<#1\right>}
\newcommand*\ichoice[1]{\mathop{\oplus}\left<#1\right>}
\newcommand*\gichoice{\mathbin{\&}}
\newcommand*\pinull{\mathbf{0}}
\newcommand*\res[1]{(\nu #1)\,}
\newcommand*\typeend{\mathtt{end}}
\newcommand*\send[1]{\mathop{\wideexclam}\left[#1\right]}
\newcommand*\rcv[1]{\mathop{\widequestion}\left[#1\right]}
\newcommand*\repl{\mathop{*}}

\newcommand*\bigparcomp\prod
\newcommand*\biggichoice\&
\newcommand*\dual[1]{\overline #1}

\newcommand*{\mktype}[1]{\mathsf{#1}}
\newcommand*{\Int}{\mktype{int}}

\newcommand*{\Bool}{\mktype{bool}}
\newcommand*{\End}{\mktype{end}}

\renewcommand*\dagger\S
\providecommand{\event}{EXPRESS/SOS 2014}

\title{Session Types as Generic Process Types}
\author{%
 Simon J.\ Gay%
\institute{School of Computing Science, University of Glasgow, UK}
\and
 Nils Gesbert%
\institute{Grenoble INP -- Ensimag, France}
\and
 Ant\'{o}nio Ravara%
\institute{CITI and
 Dep.~de Inform\'{a}tica, FCT, Universidade Nova de Lisboa, Portugal}
}

\def\titlerunning{Session Types as Generic Process Types}
\def\authorrunning{S.~J.~Gay, N.~Gesbert, and A.~Ravara}

\date{}

\begin{document}
\maketitle

\begin{abstract}
  Behavioural type systems ensure more than the usual safety
  guarantees of static analysis. They are based on the idea of
  ``types-as-processes'', providing dedicated type algebras for
  particular properties, ranging from protocol compatibility to
  race-freedom, lock-freedom, or even responsiveness.
  Two successful, although rather different, approaches, are session
  types and process types. The former allows to specify and verify
  (distributed) communication protocols using specific type (proof)
  systems; the latter allows to infer from a system specification a
  process abstraction on which it is simpler to verify properties,
  using a generic type (proof) system.
  What is the relationship between these approaches? Can the generic
  one subsume the specific one? At what price? And can the former be
  used as a compiler for the latter? The work presented
  herein is a step towards answers to such questions.
  Concretely, we define a stepwise encoding of a $\pi$-calculus with
  sessions and session types (the system of Gay and Hole
  \cite{GaySJ:substp}) into a $\pi$-calculus with process types (the
  Generic Type System of Igarashi and Kobayashi
  \cite{IgarashiA:gentspjournal}). We encode session type
  environments, polarities (which distinguish session channels
  end-points), and labelled sums. We show forward and reverse
  operational correspondences for the encodings, as well as typing
  correspondences.
  To faithfully encode session subtyping in process types subtyping,
  one needs to add to the target language record constructors and new
  subtyping rules.
  In conclusion, 
  the programming convenience of session types as protocol
  abstractions can be combined with the simplicity and power of the
  $\pi$-calculus, taking advantage in particular of the framework
  provided by the Generic Type System.
\end{abstract}

\pagestyle{plain}
\thispagestyle{empty}

\section{Introduction}
\label{sec:intro}
This work is a contribution to the understanding of the relationship
between two different type disciplines for concurrent processes,
aiming as well at compiling one into the other (considering thus the
specific session constructors as $\pi$ macros).

\paragraph{Session types \cite{HondaK:lanptd,TakeuchiK:intblt}}%
are an increasingly popular technique for specifying and verifying protocols
in concurrent and distributed systems. In a setting of point-to-point
private-channel-based communication, the session type of a channel
describes the sequence and type of messages that can be sent on it.
For example
\[
S = \echoice{\mathsf{service}:\rcv{\Int}.\send{\Bool}.\End, \mathsf{quit}:\End}
\] 
describes the server's view of a channel on which a client can select
either $\mathsf{service}$ or $\mathsf{quit}$. In the former case, the
client then sends an integer and receives a boolean; in the latter
case, the protocol ends. From the client's viewpoint, the channel has
a dual type in which the direction of messages is reversed:
\[
\dual{S} = \ichoice{\mathsf{service}:\send{\Int}.\rcv{\Bool}.\End, \mathsf{quit}:\End}
\]
Session types provide concise specifications of protocols and allow
certain properties of protocol implementations to be verified by
static type-checking.
The theory of session types was developed in order to analyse a
particular correctness criterion for concurrent systems: that every
message is of the type expected by the receiver, and that whenever a
client selects a service, the server offers a matching service.
%

\paragraph{The generic type system}(GTS, from now on) of Igarashi and
Kobayashi \cite{IgarashiA:gentspjournal} is a different approach to
type-theoretic specifications of concurrent systems: from a single
generic type system for the
$\pi$-calculus~\cite{MPW92,sangiorgi.walker:the-pi-calculus},
inferring a generic type abstracting the behaviour of the process, it
is possible to enforce specific properties by varying certain
parameters. Their motivation is to express the common aspects of a
range of type systems, enabling much of the work of designing typing
rules and proving type soundness to be packaged into a general theory
instead of being worked out for each case.  In the generic type
system, types are abstractions of processes, so that the typing rules
display a very direct correspondence between the structure of
processes and the structure of types. There is also a subtyping
relation, which can be modified in order to obtain specific type
systems; this allows, for example, a choice of retaining or discarding
information about the order of communications. A logic is provided in
which to define an $\ok$ predicate that is interpreted both as a
desired runtime property of processes and as a correctness condition
for typings. This double interpretation allows a generic type
soundness theorem to be proved, but means that type checking becomes
more like model checking unless the specific subtyping relation can be
exploited to yield an efficient type checking algorithm.

\paragraph{GTS vs.~session types.}
There is no doubt today of the usefulness of session types: expressing
protocols as type abstractions and verifying statically their
implementations is very relevant in a society of ubiquitous
computing. Process types, instead, are somehow more ``low-level'' and
are supported by powerful type systems able of ensuring a wider range
of behavioural properties. Therefore, it would be beneficial to use
process types as an executable intermediate language for
sessions-based ones.

Kobayashi \cite[Section~10]{KobayashiN:typscp} has stated that GTS
subsumes session types, although without presenting a specific
construction or giving a precise technical meaning to the term
``subsumes''. It is clearly possible, within GTS, to define a type
system similar to session types in the sense that types specify
certain allowed sequences of messages. A somewhat different question,
which we aim to answer in the present paper, is whether a specific
existing system of session types can be reproduced within GTS.

\paragraph{Related work.}
Kobayashi's paper also defines an encoding of $\pi$-calculus with
session types into $\pi$-calculus with a linear type system
\cite{KobayashiN:linpcfull} and record types, and observes that with
this encoding, subtyping for session types \cite{GaySJ:substp} arises
from the standard subtyping rules for records. That encoding is
interesting: the target language is the $\pi$-calculus with 
additional process constructs that are a good match for the branching
and selection operators of session constructors and types.

Dardha \ea~\cite{DardhaEtal:st-revisited} show an operational and a
typing correspondence, proving the correctness of Kobayashi's
encoding, and illustrate its robustness by testing it with respect to
session subtyping, polymorphism and higher-order
communication. Concretely, the target language is the $\pi$-calculus
with boolean and variant values and with a case constructor; the types
contain linear, variant, and product types. Session type duality is
captured by passing opposite capabilities in linear channel types, and
the linearised (or linearly sequential) behaviour of a session is
enforced by passing in each communication a fresh channel where the
subsequent communications of the session should take place. Not
surprisingly, branching is encoded using the case constructor and
branch and select types are encoded using variant types. Naturally,
session subtyping coincides with variant subtyping.

Demangeon and
Honda~\cite{DemangeonHonda:fullyabstractsubtypedlinearpi} study a
subtyping theory for a $\pi$-calculus augmented with branch and select
constructors (obtaining thus a fully abstract encoding of a session
calculus). In the case of this work, the source and target languages
are quite close.

\medskip
How to encode session types in process types, using GTS, is less
obvious; the types are more general and powerful, and the process
language is simpler --- just synchronous $\pi$, \ie, has no extra
features. Our aim is two relate the two different behavioural type
disciplines as they were defined, without adding extra features to the
target language. This is relevant to the design of programming
languages for distributed systems. For example: if one wants to design
a distributed object-oriented language with static typing of
protocols, can one work directly with session types instead of
developing an object-oriented formulation of process types, and
``compile'' (features of) the language into GTS?

\paragraph{Results.}
We assume that ``GTS subsumes session types'' means defining a
translation $\tsl{\cdot}$ from processes and type environments in the
source language into GTS, satisfying as many of the (usual) following
conditions as possible: (1) the encoding function should be
compositional; (2) $\tsl{P}$ should have a similar structure to $P$
(the encoding should be uniform); (3) there should be a correspondence
in both directions between the operational semantics, ideally
$P\longrightarrow Q$ if and only if $\tsl{P}\longrightarrow\tsl{Q}$;
(4) there should be a correspondence in both directions between typing
derivations, ideally $\Gamma \vdash P$ if and only if
$\tsl{\Gamma}\triangleright\tsl{P}$; (5) type soundness for session
types follows from the GTS type soundness theorem.

Defining a translation between the languages, enjoying the above
properties, is the aim of this paper.
We start by introducing the calculi under study: the next section
presents the source and the target languages (syntax and operational
semantics), and Section~\ref{sec:types} presents the types and type
systems of both languages. Then, we address three key issues: (1)
translating the polarities in the source language: $x^{+}$ and $x^{-}$
refer to the two endpoints of channel $x$ (Section~\ref{sec:enc-pol});
(2) translating the labels used in branching and selection--- external
and internal choice (Section~\ref{sec:enc-labels}); (3) comparing the
subtyping disciplines of the source and the target languages
(Section~\ref{sec:subtyping-corr}). Finally, Section~\ref{sec:concl}
concludes the paper, summarising the achievements, contributions, and
future work.

For all encodings hold forward and reverse operational
correspondences, as well as typing correspondences. The encodings
respect the properties identified above, do not restrict neither the
source nor the target language, and the operational correspondences
are as much as possible independent of the type system.
Due to space limitations, we do not present herein the failed attempts
to define the translations with the envisaged properties, nor proofs
(which, however, are all straightforward inductions over reduction
rules and typing rules; the difficulty was in formulating the
definitions).


\section{Languages}
\label{sec:languages}
We take the source language to be the version of session types defined
by Gay and Hole \cite{GaySJ:substp}.  This language uses
$\pi$-calculus $\nu$ to establish sessions, instead of special
$\mathsf{accept}/\mathsf{request}$
primitives~\cite{HondaK:lanptd,TakeuchiK:intblt}, and does not
consider progress properties \cite{DezaniCiancagliniM:sestol}. Also,
we remove recursive types, for simplicity, and we make some changes to
the structural congruence relation, to avoid inessential differences
compared with GTS.
From now on we refer to the source language --- the polarized monadic
$\pi$-calculus --- as \emph{session processes} and to the target
language --- the polyadic $\pi$-calculus --- as \emph{generic
  processes}. The source language is monadic, for simplicity, but the
target language is polyadic by demand of the encoding. %
\footnote{The syntax and the (static and dynamic) semantics of the
  source and target languages are taken from \cite{GaySJ:substp} and
  \cite{IgarashiA:gentspjournal}.}

The languages share several common process constructors: inaction,
parallel composition, scope restriction, and replication. There are
two differences. First, in session processes, channels are decorated
with \emph{polarities}, which are absent from generic processes, and
processes only synchronise when channel names have complementary
polarities. Second, session processes have constructors for
\emph{branch}, an input-labelled external choice, and \emph{select},
to choose a branch of the choice. Generic processes instead have mixed
guarded sums (but no labels), and input and output actions are
decorated with \emph{events} taken from a countable set.  Since
these tags are only relevant for properties like deadlock-freedom,
which we do not address herein, we omit them.

\paragraph{Syntax.}
\begin{myfigure}
  Choice-free source language (synchronous monadic $\pi$-calculus with polarities):
 \begin{align*}
    \textit{Choice-Free Session Processes} \quad H,J & \bnf \pinull
    \alt (H | J) \alt \res xH \alt \repl H \alt x^p\rcv{y}.H \alt x^p\send{y^q}.H \\
    \textit{Polarities} \quad p,q & \bnf + \alt - \alt \epsilon
  \end{align*}
  Full source language: adds \textit{branching} and \textit{selection}.
  \begin{align*}
    \textit{Session Processes} \quad P,Q & \bnf \pinull \alt (P | Q)
    \alt \res xP \alt \repl P \alt x^p\rcv{y}.P
    \alt x^p\send{y^q}.P \alt x^p\branch{l_i:P_i}_{i \in I} \alt x^p\sel l.P
  \end{align*}
  Full target language (synchronous polyadic $\pi$-calculus with mixed
  guarded sums):
 \begin{align*}
    \textit{Generic Processes} \quad P & \bnf \pinull \alt (P | Q)
    \alt \res xP \alt \repl P \alt \textstyle\sum_{i \in I} G_i\\
    \textit{Guarded Processes} \quad G & \bnf x\send{\tilde y}.P \alt x\rcv{\tilde y}.P
  \end{align*}
  \caption{Syntax of the source and the target languages}
  \label{fig:syntax-calculi}
\end{myfigure}

 
Consider $x,y,z$ from a countable set $\var$ of \emph{channels}.
Assume that $I$ is a non-empty finite indexing set. The source and
target languages are inductively defined by the grammars in
Figure~\ref{fig:syntax-calculi}.
As usual, $\tilde x$ abbreviates a sequence $x_1 \cdots x_n$ and
$\nu\tilde x$ abbreviates $\nu x_1 \cdots \nu x_n$, for some $n \geq
0$.

\medskip
To illustrate the session language, we present a simple example that
we incrementally develop, using it throughout the paper to also
clarify aspects of the encodings. To have a more ``realistic''
example, we assume available in the languages basic numerical values,
boolean expressions and boolean types. These are not in the syntax
(which is minimal, following the principle of Occam's razor), but can be
straightforwardly added without affecting the encodings.

The following code, parameterised on channel $x$,
implements the service branch of the server referred to in the
introduction:
\[
\mathsf{serviceBody}(x) \bdef x\rcv{i}.x\send{i=3}.\pinull
\]
The complete server, containing code to implement both $\mathsf{service}$ and
$\mathsf{quit}$, is below (the code for $\mathsf{quit}$ is trivial).
\begin{equation}\label{proc:server}
\mathsf{server}(x) \bdef
x \branch{\mathsf{service}:\mathsf{serviceBody}(x),~~\mathsf{quit}:\pinull}
\end{equation}
A client first selects one of the two options --- in this case,
$\mathsf{service}$ --- and then follows the corresponding protocol. The
definition is parameterised on channel~$x$.
\[
\mathsf{client}(x) \bdef x\sel\mathsf{service}.x\send{3}.x\rcv{b}.\pinull
\]
In a complete system, the client and server will be instantiated with
opposite endpoints of a common channel:
\[
  \mathsf{server}(c^+) \pars \mathsf{client}(c^-)
\]
This configuration is reached by one participant --- in this example,
the client --- creating the common channel and sending one endpoint to
the other participant. The channel $z$ on which this initial communication
takes place is defined by $(\nu z)$ at the top level. 
\begin{equation}\label{proc:service}
\mathsf{system} \bdef (\nu z)((\nu
c)z^+\send{c^+}.\mathsf{client}(c^-) \pars z^-\rcv{x}.\mathsf{server}(x))
\end{equation}

\paragraph{Operational semantics.} 
Let both languages be equipped with a structural pre-order (denoted
$\preceq$), along the lines of that of the generic processes,
inductively defined by the rules in Figure~\ref{fig:struct-pre}
(Page~\pageref{fig:struct-pre}), where $P \equiv Q$ stands for $P
\preceq Q$ and $Q \preceq P$.

\begin{myfigure}
Common rules (for session and generic processes):
\[
\begin{array}{c@{\extracolsep{1cm}}c}
\AxiomC{$P \rightarrow Q$}
\rrulename{R-Res}
\UnaryInfC{$\res xP \rightarrow \res xQ$}
\DisplayProof
&
\AxiomC{$P \rightarrow Q$}
\rrulename{R-Par}
\UnaryInfC{$(P \pars R) \rightarrow (Q \pars R)$}
\DisplayProof
\\[5mm]
\multicolumn{2}{c}{
\AxiomC{$P \preceq P'$}
\AxiomC{$P' \rightarrow Q'$}
\AxiomC{$Q' \preceq Q$}
\rrulename{R-SP}
\TrinaryInfC{$P \rightarrow Q$}
\DisplayProof
}
\end{array}
\]

Consider a \emph{duality} function on polarities, defined as
follows:
$\dual+=-,\
    \dual-=+,\ \textrm{and}\
    \dual\epsilon=\epsilon$.

Channel communication in session processes:
\begin{align*}
  (x^p\rcv{z}.P \pars x^{\overline p}\send{y^q}.Q)\ \rightarrow\
  (\subst Q{y^q}{z} \pars Q)
  \quad\text{with $p,q\not=\epsilon$}\quad(\textsc{R-S-Com})
\end{align*}
Labelled communication in session processes:
\begin{align*}
  (x^p\branch{l_i:P_i}_{i \in I} \pars x^{\dual p}\sel l_k.P)\
  \rightarrow\ (P_k \pars P) ~~~\text{if $k \in I$ and with
    $p\not=\epsilon$}
\quad(\textsc{R-S-ComLab})
\end{align*}
Communication in generic processes:
$$((\cdots+x\rcv{\tilde z}.P+\cdots) \pars
  (\cdots+x\send{\tilde y}.Q+\cdots))\ \to\ (\subst P{\tilde y}{\tilde
    z} \pars Q)\ \quad(\textsc{R-G-Comm})
$$
\caption{Reduction rules for processes}
\label{fig:reduction-calculi}
\end{myfigure}


The computational mechanism of the languages 
is a reduction relation on processes, inductively defined by the rules
in Figure~\ref{fig:reduction-calculi}.  The axioms use
\emph{substitution} of polarized names for unpolarized names in
processes. The definition is standard, renaming bound variables if
necessary in order to avoid capture: $\subst{P}{y^p}{x}$ denotes the
substitution of $y^p$ for the free occurrences of $x$ in $P$, and
$\subst P{\widetilde{y^p}}{\tilde x}$ denotes the simultaneous
substitution of the polarized channels in $\widetilde{y^p}$ for the
respective free occurrences of the channels in $\tilde x$ in $P$
(assuming $|\tilde x| = |\widetilde{y^p}|$, where $|\tilde x|$ denotes the
length of the sequence $\tilde x$ of channels). Recall that the
operators of the languages which introduce bindings are restriction
and the input actions. Let $\fn P$ denote the set of channels
occurring free in $P$.


\section{Types and Type Systems}\label{sec:types}
\begin{myfigure}
  \begin{align*}
    \textit{Session Types} \quad S & \bnf \mathtt{end}
    \alt \rcv{S_1}.S_2 \alt \send{S_1}.S_2 \alt \echoice{l_i:S_i}_{i \in I}
    \alt \ichoice{l_i:S_i}_{i \in I}\\
    \textit{Typing Environments} \quad \Delta & \bnf x_1^{p_1} : S_1 ;
    \ldots ; x_n^{p_n} : S_n,\,n\geq 0
  \end{align*}
  Assume that $I$ is a non-empty finite indexing set. Consider a
  countable set of labels  $l,m,l_1$, etc, disjoint from the set of
  channels.
  In a branch type $\echoice{l_i:S_i}_{i \in I}$ and in a select
  type $\ichoice{l_i:S_i}_{i \in I}$, consider all labels pairwise disjoint.
  Let $\tlbs S$ denote the set of all labels occuring at top level in
  a session type: $\tlbs{\echoice{l_i : S_i}_{i\in I}} =
  \tlbs{\ichoice{l_i : S_i}_{i\in I}} = \set{l_i | i\in I}$ and $\tlbs
  S = \0$ in the remaining cases.\\

  Consider a \emph{duality} function on session types, defined as
  follows:
  $$
    \dual\typeend=\typeend,\ \
    \dual{{\rcv{S_2}.S_1}}=\send{S_2}.\dual{S_1},\ \
    \dual{{\send{S_2}.S_1}}=\rcv{S_2}.\dual{S_1},
  $$
  $$
    \dual{{\echoice{l_i : S_i}_{i\in I}}} =\ichoice{l_i : \dual{S_i}}_{i\in I},\ \
    \textrm{and}\ \ \dual{{\ichoice{l_i : S_i}_{i\in I}}} =\echoice{l_i
      : \dual{S_i}}_{i\in I}
  $$
  A typing environment is a mapping from polarised channels into
  session types. A balanced typing environment requires
  $\Delta(x^+)=\overline{\Delta(x^-)}$ whenever $\set{x^+, x^-}
  \subseteq \dom\Delta$.
  The following rule defines a transition relation on balanced typing
  environments.
  $$\Delta,x^p:\echoice{l_i : S_i}_{i\in I},x^{\dual p}:\ichoice{l_i : S'_i}_{i\in I}\
  \xrightarrow{x,l_k}\ \Delta,x^p:S_k,x^{\dual p}:S'_k ~~~\text{if $k
    \in I$}~~~\text{\scshape (RST-Comm)}
  $$
  Finally, the rules below inductively define the type system of the
  source language.

  \centering\setlength\lineskip{5pt}
  \AxiomC{$\forall x^p\in\mathrm{dom}(\Delta), \Delta(x^p) =
    \typeend$}
  \rrulename{T-Nil}
  \UnaryInfC{$\Delta\vdash\pinull$}
  \DisplayProof\hfil
  \AxiomC{$\vdash P$}
  \rrulename{T-Rep}
  \UnaryInfC{$\vdash\repl P$}
  \DisplayProof\hfil
  \AxiomC{$\Delta_1\vdash P_1$}
  \AxiomC{$\Delta_2\vdash P_2$}
  \rrulename{T-Par}
  \BinaryInfC{$(\Delta_1 + \Delta_2)\vdash (P_1|P_2)$}
  \DisplayProof\hfil
  \AxiomC{$\Delta, x^+ : S, x^- : \overline S\vdash P$}
  \rrulename{T-New}
  \UnaryInfC{$\Delta\vdash(\nu x)\,P$}
  \DisplayProof\hfil
  \AxiomC{$\Delta, x^p : S_1, y : S_2\vdash P$}
  \rrulename{T-In}
  \UnaryInfC{$\Delta, x^p : \rcv{S_2}.S_1 \vdash x^p\rcv y.P$}
  \DisplayProof\hfil
  \AxiomC{$\Delta, x^p : S_1\vdash P$}
  \rrulename{T-Out}
  \UnaryInfC{$((\Delta, x^p : \send{S_2}.S_1) + y^q : S_2) \vdash
    x^p\send{y^q}.P$} 
  \DisplayProof\hfil
  \AxiomC{$\forall i\in I, (\Delta, x^p : S_i\vdash P_i)$}
  \rrulename{T-Offer}
  \UnaryInfC{$\Delta,x^p:\echoice{l_i : S_i}_{i\in
      I}\vdash x^p\branch{l_i:P_i}_{i\in I}$}
  \DisplayProof\hfil
  \AxiomC{$k\in I$} \AxiomC{$l = l_k$}
  \AxiomC{$\Delta, x^p : S_k\vdash P$}
  \rrulename{T-Choose}
  \TrinaryInfC{$\Delta, x^p:\ichoice{l_i : S_i}_{i\in I}\vdash x^p\sel l.P$}
  \DisplayProof
  \caption{Types and typing rules for session processes}
  \label{fig:session-types}
\end{myfigure}


\begin{myfigure}
  \begin{align*}
    \textit{Process Types} \quad \Gamma & \bnf \pinull \alt \repl\Gamma
    \alt (\Gamma_1 | \Gamma_2) \alt (\Gamma_1 \gichoice \Gamma_2)
    \alt \textstyle\sum_{i \in I} \gamma_i\\
    \textit{Guarded Types} \quad \gamma & \bnf x\send{\tau}.\Gamma
    \alt x\rcv{\tau}.\Gamma \alt \taumove.\Gamma \\
    \textit{Tuple Types} \quad \tau & \bnf (\tilde x)\Gamma
  \end{align*}
  Assume that $I$ is a non-empty finite indexing set.
  The tuple type binds the channels (variables) $\tilde x$ in
  $\Gamma$. The definition of the set of free channels in $\Gamma$
  (denoted $\fv\Gamma$) is then straightforward. A tuple or process
  type is \emph{closed} if it contains no free channels. A typing
  environment is a process type.
  \\~\\
  Let $S \subseteq \var$. Consider that $x{\downarrow}_S$ is $x$ if it
  occurs in $S$ and is $\taumove$ otherwise. Then the operation
  ${\downarrow}_S$ is defined homomorphically on closed tuple and
  process types, except in the following case: $((\tilde
  x)\Gamma){\downarrow}_S = (\tilde x)\Gamma{\downarrow}_{S \cup
    \tilde x}$.  Moreover, $\Gamma{\uparrow}_S =
  \Gamma{\downarrow}_{\var \setminus
    S}$.\\

Generic subtyping relation: base rules.
$$(\Gamma|\pinull)\eqgen\Gamma~\textsc{(Sub-Nil)}\qquad
(\Gamma_1|\Gamma_2)\eqgen(\Gamma_2|\Gamma_1)~\textsc{(Sub-Comm)}\qquad
\repl\Gamma\eqgen(\Gamma|\repl\Gamma)~\textsc{(Sub-Unfold)}$$
$$(\Gamma_1|(\Gamma_2|\Gamma_3))\eqgen((\Gamma_1|\Gamma_2)|\Gamma_3)~\textsc{(Sub-Assoc)}\qquad (\Gamma_1\&\Gamma_2)\subgen\Gamma_i
(i\in\{1,2\})~\textsc{(Sub-IChoice)}$$
{\centering\setlength\lineskip{10pt}
\AxiomC{$\Gamma\subgen\Gamma'$}
\rrulename{Sub-Rep}
\UnaryInfC{$\repl\Gamma\subgen\repl\Gamma'$}
\DisplayProof
\hfil
\AxiomC{$\Gamma\subgen\Gamma'$}
\rrulename{Sub-Abs}
\UnaryInfC{$(\tilde x)\Gamma\subgen(\tilde x)\Gamma'$}
\DisplayProof
\hfil
\AxiomC{$\Gamma\subgen\Gamma'$}
\rrulename{Sub-Restrict}
\UnaryInfC{$\Gamma{\downarrow}_S\subgen\Gamma'{\downarrow}_S$}
\DisplayProof
\hfil
\text{}\\[.5em]
\AxiomC{$\Gamma\subgen\Gamma'$}
\rrulename{Sub-Subs}
\UnaryInfC{$\subst{\Gamma}{\tilde y}{\tilde x}\subgen\subst{\Gamma'}{\tilde y}{\tilde x}$}
\DisplayProof
\hfil
\AxiomC{$\Gamma_1\subgen\Gamma'_1$}\AxiomC{$\Gamma_2\subgen\Gamma'_2$}
\rrulename{Sub-Par}
\BinaryInfC{$(\Gamma_1|\Gamma_2)\subgen(\Gamma'_1|\Gamma'_2)$}
\DisplayProof
\hfil
}\\[.5em]

Generic subtyping relation: additional rules.
$$\repl\pinull\eqgen\pinull~\textsc{(Sub-Inact)}\ \qquad\
    \Gamma\eqgen\taumove.\Gamma~\textsc{(Sub-TPref)}\ \qquad\
    \Gamma{\downarrow}_V|\Gamma{\uparrow}_V\subgen\Gamma~\textsc{(Sub-Divide)}
$$
  \null\hfil\AxiomC{$\forall i\in I, \gamma_i \subgen \gamma'_i$}
  \rrulename{Sub-Choice}
  \UnaryInfC{$\sum_{i\in I}\gamma_i \subgen \sum_{i\in I}\gamma'_i$}
  \DisplayProof\\[.5em]

Generic typing rules:

\centering\setlength\lineskip{10pt}
\AxiomC{}
\rrulename{T-Nil}
\UnaryInfC{$\pinull\triangleright\pinull$}
\DisplayProof\hfil
\AxiomC{$\Gamma\triangleright P$}
\rrulename{T-Rep}
\UnaryInfC{$\repl\Gamma\triangleright \repl P$}
\DisplayProof\hfil
\AxiomC{$\Gamma_1\triangleright P_1$}\AxiomC{$\Gamma_2\triangleright P_2$}
\rrulename{T-Par}
\BinaryInfC{$(\Gamma_1 | \Gamma_2)\triangleright (P_1 | P_2)$}
\DisplayProof\hfil
\AxiomC{$\Gamma'\triangleright P$}\AxiomC{$\Gamma\subgen\Gamma'$}
\rrulename{T-Sub}
\BinaryInfC{$\Gamma\triangleright P$}
\DisplayProof\hfil
\AxiomC{$\Gamma_1|\Gamma_2\triangleright P$}
\AxiomC{$\{\tilde y\} \cap \fv{\Gamma_1}=\emptyset$}
\rrulename{T-In}
\BinaryInfC{$x\rcv{(\tilde y)\Gamma_2}.\Gamma_1\triangleright x\rcv{\tilde y}.P$}
\DisplayProof\hfil
\AxiomC{$\Gamma_1\triangleright P$}
\rrulename{T-Out}
\UnaryInfC{$x\send{(\tilde y)\Gamma_2}.(\Gamma_1|\subst {\Gamma_2}{\tilde z}{\tilde y})
  \triangleright x\send{\tilde z}.P$}
\DisplayProof\hfil
  \AxiomC{$\forall i\in I, \gamma_i\triangleright G_i$}
  \rrulename{T-Choice}
  \UnaryInfC{$\sum_{i\in I}\gamma_i\triangleright\sum_{i\in I}G_i$}
  \DisplayProof\hfil
\AxiomC{$\Gamma\triangleright P$}
\AxiomC{$\ok(\Gamma{\downarrow}_{\tilde x})$}
\AxiomC{$\fv{\Gamma{\uparrow}_{\tilde x}} \cap \{\tilde x\}=\emptyset$}
\rrulename{T-New}
\TrinaryInfC{$\Gamma{\uparrow}_{\tilde x}\triangleright (\nu \tilde x)\,P$}
\DisplayProof
\caption{Types, subtyping and typing for generic processes}
\label{fig:gen-types}
\end{myfigure}


\begin{myfigure}
\centering\setlength\lineskip{5pt}
\AxiomC{$\tau_1\subgen\tau_2$}
\rrulename{RPT-Comm}
\UnaryInfC{$(x\send{\tau_1}.\Gamma_1 | x\rcv{\tau_2}.\Gamma_2)\ \to\ (\Gamma_1|\Gamma_2)$}
\DisplayProof
\hfil
%
%
$\taumove.\Gamma \to \Gamma$~{\scshape (RPT-Evt)}
\\
\AxiomC{$\Gamma\to\Gamma'$}
\rrulename{RPT-Par}
\UnaryInfC{$(\Gamma | \Gamma_1)\ \to\ (\Gamma' | \Gamma_1)$}
\DisplayProof\hfil
\AxiomC{$\Gamma_1\subgen\Gamma'_1$}
\AxiomC{$\Gamma'_1\to\Gamma'_2$}
\AxiomC{$\Gamma'_2\subgen\Gamma_2$}
\rrulename{RPT-Sub}
\TrinaryInfC{$(\Gamma_1|\Gamma_2)\ \subgen\ (\Gamma'_1|\Gamma'_2)$}
\DisplayProof
\caption{Reduction on process types}
\label{fig:red-proc-types}
\end{myfigure}


We define now the type systems of both languages, and summarise the
results for each system.

\paragraph{Session types.}
Figure~\ref{fig:session-types} defines the syntax of session types and
type environments, and the corresponding typing rules.  Several rules
use the $+$ operation on type environments
(Definition~\ref{def:unG}). This operation is defined for a type
environment and a typed identifier, and then extended inductively. It
is a partial operation, and if it occurs in a rule then definedness of
the operation is an implicit hypothesis of the rule.

\begin{definition}[from \cite{GaySJ:substp}]\label{def:unG}
  \begin{enumerate}
  \item Let $\Delta,x^p:S=\Delta$, if $x^p:S \in \Delta$, and
    $\Delta,x^p:S=\Delta \cup \set{x^p:S}$, if $x^p \notin
    \dom\Delta$. Otherwise the operation is undefined.
  \item Consider $p \not= \epsilon$. Let $\Delta+x^p:S=\Delta,x^p:S$,
    if $\set{x^p,x^\epsilon} \cap \dom\Delta = \0$; let
    $\Delta+x^\epsilon:S=\Delta,x^\epsilon:S$, if
    $\set{x^+,x^-,x^\epsilon} \cap \dom\Delta = \0$. Otherwise the
    operation is undefined.
  \end{enumerate}
\end{definition}
One easily concludes that the system presented in
Section~\ref{sec:languages} as Process (\ref{proc:service}) is
well-typed: $\vdash \mathsf{system}$.
More interesting is the typing of the process below the restrictions:
\[
z^+:\send{S}.\typeend, z^-:\rcv{S}.\typeend, c^+:S, c^-:\dual{S} \vdash
z^+\send{c^+}.\mathsf{server}(c^+) \pars z^-\rcv{x}.\mathsf{client}(x)
\]
where the session types $S$ and $\dual{S}$ are defined in
Section~\ref{sec:intro}. 

\medskip
Gay and Hole \cite{GaySJ:substp} proved Type Preservation and Type
Safety (no errors in well-typed processes).
Appendix~\ref{app:results-languages} presents the results adapted to the
setting we use herein (Theorems \ref{thm:tp} and \ref{thm:sf}).

\paragraph{Generic types.}
GTS is parameterized by a subtyping relation and a consistency
condition on types: instantiating them yields a particular type
system, ensuring a given (safety) property on processes. These
``generic'' conditions occur in the typing rules. Type soundness
results depend on the particular semantic property on processes one is
interested in. We present thus the particular subtyping relation and
consistency predicate used in this work:
Figure~\ref{fig:gen-types} defines the syntax of generic process
types, and presents the rules defining the subtyping and the typing
relations.
Most constructors are fairly standard in process algebras: `*' stands
for replication, `\&' for internal choice, `+' for external choice,
and `$\taumove$' for a synchonisation event.
The subtyping relation $\subgen$ is the preorder satisfying
the rules presented in the figure, where $\Gamma_1 \eqgen \Gamma_2$ if
$\Gamma_1 \subgen \Gamma_2$ and $\Gamma_2 \subgen \Gamma_1$.

Since types are themselves processes, one needs to define an
operational semantics to describe their evolution. Let
$\subst{\Gamma}{\tilde y}{\tilde x}$ denote the simultaneous
capture-avoiding substitution of the channels in $\tilde y$ for the
respective free occurrences of the channels in $\tilde x$ in
$\Gamma$. Reduction on process types is inductively defined by the
rules in Figure~\ref{fig:red-proc-types}. Write $\to^*$ for the
reflexive and transitive closure of $\to$.

The process types that ``behave correctly'' are those that are
\emph{well-formed}.

\begin{definition}[from \cite{IgarashiA:gentspjournal}]\label{def:wf}
  A process type $\Gamma$ is \emph{well-formed}, written $\wfg$, if
  whenever $\Gamma \to^* (x\send{\tau_1}.\Gamma_1 |
  x\rcv{\tau_2}.\Gamma_2 | \Gamma_3)$ then $\tau_1 \subgen \tau_2$
  holds.
\end{definition}

To ensure particular conditions on the behaviour of process types, GTS
introduces the notion of \emph{consistency predicate}. It uses the
auxiliary notion of non-active process type, in the form of a
$\opnsf{NULL}$ predicate.

\begin{definition}[from \cite{IgarashiA:gentspjournal}]\label{def:ok}
  Let $\opnsf{NULL}(\Gamma)$ hold if $\Gamma$ has no subprocess input
  or output guarded.  A predicate \ok\ on process types is a
  \emph{proper consistency predicate} if it is preserved by reduction,
  if $\okg$ then $\wfg$, and if $\okg$ and $\opnsf{NULL}(\Gamma')$
  then $\ok(\Gamma|\Gamma')$.
\end{definition}

The particular consistency condition we are interested in is
\emph{linearity}: a process type has no parallel sends or receives on
a given channel.

\begin{definition}\label{def:lin}
  A process type $\Gamma$ is \emph{linear}, written $\ling$, if
  $\wfg$, if $\Gamma \to^* (x\send{\tau_1}.\Gamma_1 | \Gamma_2)$ implies
  $\Gamma_2 \not\to^* (x\send{\tau_2}.\Gamma_3 | \Gamma_4)$, and if
  $\Gamma \to^* (x\rcv{\tau_1}.\Gamma_1 | \Gamma_2)$ implies $\Gamma_2
  \not\to^* (x\rcv{\tau_2}.\Gamma_3 | \Gamma_4)$.
\end{definition}

\begin{lemma}\label{lm:lin-pcp}
  $\ling$ is a proper consistency predicate.
\end{lemma}

Igarashi and Kobayashi proved Subject Reduction --- as processes
evolve ``their'' (well-formed) process types evolve accordingly ---
and, as a corollary, a property preservation result relating
properties on processes and on process types: let $p$ be an invariant
predicate on processes and consider an \ok\ predicate that is its
correspondent on process types; a consequence of Subject Reduction is
that, if a process type typing a given process satisfies an \ok\
predicate, the process satisfies the invariant $p$ (Theorem \ref{thm:ts}
in Appendix~\ref{app:results-languages}).
Hereafter, $\ling$ is our \ok\ predicate.

The counterpart of an error session process is a generic process with
an arity mismatch or with races (parallel sends or receives on the same
channel).

\begin{definition}[Error process]\label{def:errGP}
  A generic process $P$ is \emph{an error}, if one of the three
  conditions below hold.
  \begin{enumerate}
  \item Whenever $P \equiv \res{\tilde x}(x\rcv{\tilde z}.P_1 \pars
    x\send{\tilde y}.P_2|Q)$ it is the case that $|\tilde z| \not=
    |\tilde y|$;
  \item $P \equiv \res{\tilde x}(x\send{\tilde z}.P_1 \pars
    x\send{\tilde y}.P_2|Q)$;
  \item $P \equiv \res{\tilde x}(x\rcv{\tilde z}.P_1 \pars
    x\rcv{\tilde y}.P_2|Q)$. 
  \end{enumerate}
\end{definition}

The counterpart of session type safety is simply linearity and the
absence of arity mismatches. Linearity implies the absence of races,
as ensured usually by session types; the definition of well-formedness
(Definition~\ref{def:wf} above) implies no arity mismatches, allowing
however a channel to change its arity after a reduction step. The
predicate \linop\ states these conditions on process types: in the
original process and after each reduction step, there are no parallel
sends or receives on the same channel nor arity mismatches (this last
condition resulting from \linop\ being a consistency predicate ---
Lemma~\ref{lm:lin-pcp}).

To ensure that well-typed generic processes are type safe, it suffices
to show that absence of errors in generic processes (an invariant
property) corresponds to \linop. The result is a corollary of a
theorem proved by Igarashi and Kobayashi.

\begin{proposition}[Resulting from Theorem 5.1 of
  \cite{IgarashiA:gentspjournal}]\label{thm:cons-cf}
  If $\Gamma\triangleright P$ and $\opnsf{LIN}(\Gamma)$ then $P$ is not an error.
\end{proposition}


\section{Encoding Polarities}\label{sec:enc-pol}
\begin{myfigure}
  Let $\phi$ be a name translation function.
  The translation of processes is defined by the following axioms
  (here $p$, $q$ may be $+$, $-$ or $\epsilon$), assuming the names
  $u$ and $v$ fresh, being homomorphic in the cases omitted.
  \begin{align*}
    \tsl{x^p\send{y^q}.P}_\phi & =
    \tsl{x^p\send{y^q}}_\phi.\tsl{P}_\phi &
    \tsl{x^p\rcv{y}.P}_\phi & =
    \tsl{x^p\rcv{y}}^{(u,v)}_\phi.\tsl{P}_{\phi+\{y\mapsto(u,v)\}} \\
    \tsl{(\nu x)\,P}_\phi & = (\nu
    u,v)\,\tsl{P}_{\phi+\{x\mapsto(u,v)\}}
  \end{align*}
  The translation of prefixes is defined as follows, where
  $p,q\not=\epsilon$.
  \begin{align*}
    \tsl{x^p\send{y^q}}_\phi & =
    \phi^{\overline{p}}(x)\send{\phi^q(y),\phi^{\overline{q}}(y)} &
    \tsl{x^p\send{y}}_\phi & =
    \phi^{\overline{p}}(x)\send{\phi^+(y),\phi^-(y)}\\
    \tsl{x\send{y^q}}_\phi & =
    \phi^-(x)\send{\phi^q(y),\phi^{\overline{q}}(y)} &
    \tsl{x\send{y}}_\phi & =
    \phi^-(x)\send{\phi^+(y),\phi^-(y)}\\
    \tsl{x^p\rcv{y}}^{(u,v)}_\phi & =
    \phi^p(x)\rcv{u,v} &
    \tsl{x\rcv{y}}^{(u,v)}_\phi & =
    \phi^+(x)\rcv{u,v}
  \end{align*}
  \caption{Process translation}
  \label{fig:proc-transl-def}
\end{myfigure}


The first question in defining a translation from session processes to
generic processes is how to represent polarities. It is easy to see
that a non-trivial encoding is necessary. Simply erasing polarities
would lead to reductions in the target language that are not possible
in the source language; this is clear from \textsc{R-S-Com} and
\textsc{R-G-Com}.
We do not restrict the encoding to well-typed processes, as we want to
show that it is possible to define a translation where the session
types system and an instance of GTS yield the same classification of
typable processes.

The translation guarantees an operational and a typing
correspondence. To state the former in the reverse direction, we need
to restrict the result to well-typed processes.

To simplify the presentation in this section we consider only
choice-free (\ie, non-branching, using only sequence and parallel)
processes and types.

\paragraph{The process translation} function maps each free name of
the process into two new target names, according to the rules in
Figure~\ref{fig:proc-transl-def}. The translation uses a \emph{name
  translation} (partial) \emph{function} $\phi$ from names to pairs of
names. Write $\phi^+$ and $\phi^-$ for the compositions of $\phi$ with
the first and second projections, respectively. We require $\phi^+$ and
$\phi^-$ to be injective and have disjoint images. When $\dom{\phi}$
and $\dom{\psi}$ are disjoint, we write $\phi+\psi$ to denote the
union of the name translation functions $\phi$ and $\psi$
(otherwise, this operation is undefined).

To illustrate this encoding, consider the process $\mathsf{system}$
(Process \ref{proc:service} in Page \pageref{proc:service}).
The encoding of the process is below, considering $\phi
\bdef \set{x \mapsto (t,u)}, \chi \bdef \set{z \mapsto (v, w)}, \psi
\bdef \set{c\mapsto (d, e)}$. 
\begin{eqnarray*}\label{proc:serviceEnc}
  \tsl{\mathsf{system}}_\emptyset & = &
  \res{v,w}\tsl{\res c z^+\send{c^+}.\mathsf{client}(c^-) \pars
    z^-\rcv{x}.\mathsf{server}(x)}_{\chi}\\ & = &
  \res{v,w}(\res{d,e}w\send{d, e}.\tsl{\mathsf{client}(c^-)}_{\chi+\psi})
  \pars w\rcv{t, u}.\tsl{\mathsf{server}(x)}_{\phi+\chi}
%
\end{eqnarray*}
Thus, the communication between $z^+$ and $z^-$ becomes a
communication on $w$.
\paragraph{The forward operational correspondence}
between source and target reduction steps is one-to-one.  Let $P$ be a
choice-free session process and $\phi$ be a name translation function.
%
%
\begin{lemma}\label{lm:sub-cf}
  \begin{enumerate}
  \item If $p\not=\epsilon$ then $\tsl{\subst{P}{x^p}{y}}_\phi =
    \tsl{P}_{\phi+\{y\mapsto(\phi^p(x),\phi^{\overline{p}}(x))\}}$
  \item $\tsl{\subst{P}{x}{y}}_\phi =
    \tsl{P}_{\phi+\{y\mapsto(\phi^+(x),\phi^-(x))\}}$
  \end{enumerate}
\end{lemma}

\begin{theorem}\label{thm:foc-cf}
  If $P\longrightarrow Q$ then $\tsl{P}_\phi\longrightarrow\tsl{Q}_\phi$.
\end{theorem}

\begin{myfigure}
  Let $\phi$ be the mapping from names to pairs of names, introduced
  in Figure~\ref{fig:proc-transl-def}. 

  Consider $\tsls{S} = (u,v)\tsl{y:S}_{\{y\mapsto(u,v)\}\cup\varphi}$,
  where $u$, $v$, and $y$ are fresh.
  $$\tsl{\set{x_1^{p_1}:S_1,\ldots,x_n^{p_n}:S_n}}_\phi =
     \tsl{x_1^{p_1}:S_1}_\phi \pars \cdots \pars \tsl{x_n^{p_n}:S_n}_\phi$$
\null\vskip -6ex
  \begin{align*}
    \tsl{x^p:\mathtt{end}}_\phi & = \pinull\\
    \tsl{x:\send{S_1}.S_2}_\phi & =
     \phi^-(x)\send{\tsls{S_1}}.\tsl{x:S_2}_\phi &
    \tsl{x:\rcv{S_1}.S_2}_\phi & =
     \phi^+(x)\rcv{\tsls{S_1}}.\tsl{x:S_2}_\phi\\
    \tsl{x^+:\rcv{S_1}.S_2}_\phi & =
     \phi^+(x)\rcv{\tsls{S_1}}.\tsl{x^+:S_2}_\phi &
    \tsl{x^-:\rcv{S_1}.S_2}_\phi & =
     \phi^-(x)\rcv{\tsls{S_1}}.\tsl{x^-:S_2}_\phi\\
    \tsl{x^+:\send{S_1}.S_2}_\phi & =
     \phi^-(x)\send{\tsls{S_1}}.\tsl{x^+:S_2}_\phi &
    \tsl{x^-:\send{S_1}.S_2}_\phi & =
     \phi^+(x)\send{\tsls{S_1}}.\tsl{x^-:S_2}_\phi
  \end{align*}
  \caption{Type environment translation}
  \label{fig:typeenv-seq-transl-def}
\end{myfigure}


\paragraph{The type translation}(in
Figure~\ref{fig:typeenv-seq-transl-def}) maps sequential session types
to generic (choice-free) process types. Let us first present an
example.
Considering
$$\Delta \bdef
  \set{c^+:\rcv{\Int}.\send{\Bool}.\End,c^-:\send{\Int}.\rcv{\Bool}.\End}
$$
we have $\Delta \vdash c\rcv{i}.c\send{i=3}.\pinull \pars c\send{3}.c\rcv{b}.\pinull$.
Then,
\begin{eqnarray*}\label{type:serviceEnc}
  \Gamma \bdef \tsl{\Delta}_\psi & = &
    \tsl{\set{c^+:\rcv{\Int}.\send{\Bool}.\End}}_\psi \pars
    \tsl{\set{c^-:\send{\Int}.\rcv{\Bool}.\End}}_\psi \\ & = &
    d\rcv{\Int}.e\send{\Bool}.\pinull \pars d\send{\Int}.e\rcv{\Bool}.\pinull
\end{eqnarray*}
It is easy to check that the encoded system is typable with the
encoding of the type environment (\ie, $\Gamma \triangleright
\tsl{\mathsf{system}}_\psi$). However, the generic process type no
longer captures the flow of the protocol, as the two steps (exchanging
first an integer and then a boolean) happen now on different channels.
If one thinks of a process with (possibly long and complex) sessions,
one understands that the encoding produces a large number of new
channels, requiring a partial order on them to exibit the flow of the
protocol that is clear in each session type.



\paragraph{The reverse operational correspondence}
requires typing: a communication between, for example, send
on $x^-$ and receive on $x$, does not reduce in the source language
(and is ill-typed) but translates into a reduction in the target
language.

\begin{theorem}\label{thm:roc-cf}
  Let $P$ be a well-typed choice-free session process and let $\phi$
  be a name translation function. If $\tsl{P}_\phi \longrightarrow Q$
  then there exists $P'$ such that $P\longrightarrow P'$ and
  $\tsl{P'}_\phi = Q$.
\end{theorem}

\paragraph{Typing correspondence.}  We show a correspondence in both
directions between typing derivations. Let $P$ be a choice-free
session process and let $\phi$ be a name translation function. We
state first completeness and then soundness.

\begin{theorem}\label{thm:compl-cf}
  If $\Delta\vdash P$ for some balanced $\Delta$, then
  $\tsl{\Delta}_\phi\triangleright\tsl{P}_\phi$ and
  $\opnsf{LIN}(\tsl{\Delta}_\phi)$.
\end{theorem}

The converse of completeness does not hold. For example, take $P =
\repl\pinull$. Then $\tsl{x^+:\End}\triangleright\tsl{\repl\pinull}$,
but it is not the case that $x^+:\End\vdash\repl\pinull$.

\begin{theorem}\label{thm:sound-cf}
  If $\Gamma\triangleright\tsl{P}_\phi$ and $\opnsf{LIN}(\Gamma)$ then
  $\Delta\vdash P$, for some balanced $\Delta$.
\end{theorem}

Note that $\Gamma$ must be linear, otherwise, considering $\phi(x) =
(u,v)$, we have $(u\send{} | u\send{})\ \triangleright\
(u\send{} \pars u\send{})$, whereas $(x^+\send{} \pars x^+\send{})$ is
not typable as a session process.


\section{Encoding labels}\label{sec:enc-labels}
Figure~\ref{fig:proc-transl-branch-def} extends the translation of
processes to include labelled sums (branch) and selectors. We
translate the labels occuring in a branch process as fresh names that
are sent to the translation of the corresponding select process; the
latter, in turn, selects its desired branch outputing on the name
corresponding to the label. We use a function $\sigma$ to map the
labels to the fresh channels.

To translate the select process, we need typing information (an
environment typing the process) to know how many labels the
corresponding branch has. That is the number of parameters of the
channel which is waiting for the fresh names created by the other
end-point to represent the labels.
\begin{myfigure}
Let $\Delta$ be a typing environment.\\

\noindent
Extend the translation of prefixes with the following rules, where $p\not=\epsilon$.
\begin{align*}
\tsl{x^p\sel l}^\Delta_\phi & =
\phi^p(x)\rcv{\lambda_1\ldots\lambda_n}\textrm{, where }
  \set{\lambda_1\ldots\lambda_n} = \tlbs{\Delta(x^{\dual{p}})}\textrm{, with }n \geq 0 \\
\tsl{x\sel l}^\Delta_\phi & =
\phi^+(x)\rcv{\lambda_1\ldots\lambda_n}\textrm{, where }
  \set{\lambda_1\ldots\lambda_n} = \tlbs{\Delta(x^{\dual{p}})}\textrm{, with }n \geq 0
\end{align*}

Extend the translation of processes with the following rules, where
$p \in \set{+,-,\epsilon}$. For all $i \in \set{1,\ldots,n}$ let
$\set{\lambda_1\ldots\lambda_n}\cap\fn{P_i }=\0$, and
let $\Delta \xrightarrow{x,l_i} \Delta_i$. Let $\sigma(l)$ = $\lambda_j$ where $\Delta(x^p) = \ichoice{l_i:S_i}_{i=1}^n$ and $l = l_j$.
%
  \begin{align*}
  \tsl{x^p\sel l.P}^\Delta_\phi & =
  \tsl{x^p\sel l}^\Delta_\phi.\sigma(l)\send{}.\tsl{P}^{\Delta_i}_\phi\\
  \tsl{x^+\branch{l_i : P_i}_{i=1}^n}^\Delta_\phi & =
   (\nu\lambda_1\ldots\lambda_n)\,\phi^-(x)\send{\smash{\lambda_1\ldots\lambda_n}}.
   \textstyle\sum_{i=1}^n\lambda_i\rcv{}.\tsl{P_i}^{\Delta_i}_\phi\\
  \tsl{x^-\branch{l_i : P_i}_{i=1}^n}^\Delta_\phi & =
   (\nu\lambda_1\ldots\lambda_n)\,\phi^+(x)\send{\smash{\lambda_1\ldots\lambda_n}}.
   \textstyle\sum_{i=1}^n\lambda_i\rcv{}.\tsl{P_i}^{\Delta_i}_\phi\\
  \tsl{x\branch{l_i : P_i}_{i=1}^n}^\Delta_\phi & =
   (\nu\lambda_1\ldots\lambda_n)\,\phi^-(x)\send{\smash{\lambda_1\ldots\lambda_n}}.
   \textstyle\sum_{i=1}^n\lambda_i\rcv{}.\tsl{P_i}^{\Delta_i}_\phi
  \end{align*}
  \caption{Process translation for branch and select}
  \label{fig:proc-transl-branch-def}
\end{myfigure}


\medskip
The illustrate the idea, we encode now the $\mathsf{server}$
(process~\ref{proc:server}). Recall that $\phi \bdef \set{x \mapsto
  (t,u)}$ and $\chi \bdef \set{z \mapsto (v, w)}$,
$S = \echoice{\mathsf{service}:\rcv{\Int}.\send{\Bool}.\End,
  \mathsf{quit}:\End}$, and

\noindent $\Delta \bdef
\set{x^+:\rcv{\Int}.\send{\Bool}.\End,x^-:\send{\Int}.\rcv{\Bool}.\End}$.
Consider now $\Delta' \bdef \set{x^+:S,x^-:\dual S}$ and let the
function $\sigma$ associate the labels \textit{service} and
\textit{quit} with the channels \textit{service} and \textit{quit}. Then,
\begin{eqnarray*}\label{proc:serverEnc}
  \tsl{\mathsf{server}(x^+)}^{\Delta'}_{\phi+\chi} & = &
    \res{\mathit{service},\mathit{quit}} u\send{\mathit{service},\mathit{quit}}.\\ &  &
     (\mathit{service}\rcv{}.\tsl{\mathsf{serviceBody}(x^+)}^{\Delta'}_{\phi+\chi} + 
      \mathit{quit}\rcv{}.\pinull)
\end{eqnarray*}

\paragraph{Forward operational correspondence.}
We no longer have a one-to-one correspondence between reduction steps,
because communication on a label in the source language is translated
into two communications.

\begin{theorem}
  Whenever $P\longrightarrow Q$, then
  $\tsl{P}\longrightarrow^n\tsl{Q}$ with $n = 1$ or $n = 2$.
\end{theorem}

\paragraph{Reverse operational correspondence.}
This must also take into account the extra reductions involved in
communication of labels. The second case in the theorem describes the
intermediate configuration between the two reduction steps
corresponding to communication of a label.

\begin{theorem}
  Let $\Delta \vdash P$ with $\Delta$ balanced and let $\phi$
  be a name translation function. If $\tsl{P}^\Delta_\phi \longrightarrow Q$
  then there exists $\Delta'$ and $P'$ such that $P\longrightarrow P'$ and either
  $\tsl{P'}^{\Delta'}_\phi = Q$ or $Q \equiv \res{\tilde\lambda}Q_1$
  and $Q_1 \to Q_2 \equiv \tsl{P'}^{\Delta'}_\phi$.
\end{theorem}
Notice that, by Subject-Reduction, $\Delta'$ is balanced and $\Delta' \vdash P'$.

\paragraph{Typing correspondence.}
Figure~\ref{fig:typeenv-transl-def} extends the translation of typing
environments to branching types.  Error processes in the full session
language must now take into account the possibility of a selector
``asking for'' a non-existing label in a branching offer:

if $P \equiv \res{\tilde x}(x^p\branch{l_i:P_i}_{i \in I} \pars
x^{\dual p}\sel l.Q \pars R)$ then either $l=l_k$ and $k \notin I$
or $\set{x^+, x^-} \cap \fn R \not= \0$.

\medskip
The \ok\ predicate is still simply $\opnsf{LIN}$.
Theorems~\ref{thm:compl-cf} and \ref{thm:sound-cf} have corresponding
versions for the full calculi: a session process is well-typed if and
only if its encoding is well-typed in an \ok\ typing environment.
In the following results, let $P$ be a session process and let $\phi$
be a name translation function.

\begin{theorem}[Completeness]\label{thm:compl}
  If $\Delta\vdash P$ then
  $\tsl{\Delta}_\phi\triangleright\tsl{P}^{\Delta}_\phi$ and
  $\opnsf{LIN}(\tsl{\Delta}_\phi)$.
\end{theorem}

\begin{theorem}[Soundness]\label{thm:sound}
  Let $\Delta'$ be a balanced session type environment. If
  $\Gamma\triangleright\tsl{P}^{\Delta'}_\phi$ and $\opnsf{LIN}(\Gamma)$ then
  $\Delta\vdash P$, for some balanced $\Delta$.
\end{theorem}

\begin{myfigure}
  Let $\phi$ be the mapping from names to pairs of names, introduced
  in Figure~\ref{fig:proc-transl-def}.

  Consider $\tsls{S} = (u,v)\tsl{y:S}_{\{y\mapsto(u,v)\}\cup\varphi}$,
  where $u$, $v$, and $y$ are fresh.

  For all $i \in \set{1,\ldots,n}$ let
  $\set{\lambda_1\ldots\lambda_n}\cap\fn{P_i }=\0$.
  %
  \begin{align*}
    \tsl{x^+:\echoice{l_i:S_i}_{i=1}^n}_\phi & =
    \phi^-(x)\send{\lambda_1,\ldots,\lambda_n}\textstyle\sum_{i=1}^n
    \lambda_i\rcv{\tsls{S_i}} .\pinull \\
    \tsl{x^-:\echoice{l_i:S_i}_{i=1}^n}_\phi & =
    \phi^+(x)\send{\lambda_1,\ldots,\lambda_n}\textstyle\sum_{i=1}^n
    \lambda_i\rcv{\tsls{S_i}}.\pinull \\
    \tsl{x:\echoice{l_i:S_i}_{i=1}^n}_\phi & =
    \phi^-(x)\send{\lambda_1,\ldots,\lambda_n}\textstyle\sum_{i=1}^n
    \lambda_i\rcv{\tsls{S_i}}.\pinull \\
    \tsl{x^+:\ichoice{l_i:S_i}_{i=1}^n}_\phi & =
    \phi^+(x)\rcv{\lambda_1,\ldots,\lambda_n}\textstyle\biggichoice_{i=1}^n
    \lambda_i\send{\tsls{\overline{S_i}}}.\pinull \\
    \tsl{x^-:\ichoice{l_i:S_i}_{i=1}^n}_\phi & =
    \phi^-(x)\rcv{\lambda_1,\ldots,\lambda_n}\textstyle\biggichoice_{i=1}^n
    \lambda_i\send{\tsls{\overline{S_i}}}.\pinull \\
    \tsl{x:\ichoice{l_i:S_i}_{i=1}^n}_\phi & =
    \phi^+(x)\rcv{\lambda_1,\ldots,\lambda_n}\textstyle\biggichoice_{i=1}^n
    \lambda_i\send{\tsls{\overline{S_i}}}.\pinull
  \end{align*}
  \vspace{-8mm}
  \caption{Type environment translation}
  \label{fig:typeenv-transl-def}
\end{myfigure}



\section{Subtyping correspondence}\label{sec:subtyping-corr}
Subtyping is an essential ingredient of the theory of session
types. Originally proposed by Gay and Hole~\cite{GaySJ:substp}, it has
been widely used in other session-based systems, with
subject-reduction and type-safety holding. We now discuss how to
represent session subtyping in GTS.

\paragraph{Safe substitutability.}
Notice that subtyping in session types means less branching (`\&') and
more choice (`$\oplus$'). This basic principle conforms to the ``safe
substitutability principle'' of Liskov and
Wing~\cite{liskov.wing:behavioural-subtyping}.
However, the principle has no counterpart in process types: although the
internal choice axiom follows the principle, the rule for external choice
(`+') does not allow changing the number of arguments of the
operation (Appendix~\ref{app:subtyping} presents the subtyping
relations on session types and on generic process types).

It is thus not surprising that the encoding presented in the previous
section does not preserve subtyping. Let
$S_1 \bdef \echoice{l_1:\End}$ and $S_2 \bdef \echoice{l_1:\End,l_2:\End}$;
we have $S_1 \subsess S_2$, but the encoded types are not related:
$$\tsls{S_1} =
(u,v)v\send{l_1}l_1\send{\tsls{\End}} .\pinull \not\subgen
(u,v)v\send{l_1,l_2}(l_1\send{\tsls{\End}} \& l_2\send{\tsls{\End}})
.\pinull = \tsls{S_2}$$
The problem is twofold: (1) the subtyping relation on process
types does not allow changing the sequence of names sent on an
output; (2) session subtyping means less branching, but process
subtyping on choice does not allow to change the number of summands.
%

To achieve an encoding guaranteeing a typing correspondence one would
have to ``compensate'' subtyping, always passing all labels occurring at top
level in the branch process of a given channel. Consider
$$
 \Delta \bdef \set{x^+:\echoice{l_1:\End},x^-:\ichoice{l_1:\End}}
 \textrm{ and }
 P \bdef
 (x^+\branch{\mathsf{l_1}:\pinull,\mathsf{l_2}:\pinull} \pars
  x^-\sel\mathsf{l_1}.\pinull)
$$
Note that $\Delta \vdash P$. To avoid breaking the typing
correspondence, instead of encoding $\Delta$ and $P$ independently, if
one encodes the typing judgement then the encoding of $\Delta$ needs
to take into account all the labels of the branch offered by $x^+$. It
is easy to get an encoding preserving and reflecting typability, but
actually, the idea presented removes subtyping. The interesting
question is whether an encoding in GTS, capturing the
session type subtyping discipline, exists or not.

\paragraph{Variant subtyping.}
A straightforward way of faithfully representing session subtyping in
GTS is to extend the language of processes with labelled values and a
case constructor, and the language of types with variants (as done by
Dardha \ea~\cite{DardhaEtal:st-revisited} and by Demangeon and
Honda~\cite{DemangeonHonda:fullyabstractsubtypedlinearpi}). Using the usual rules of
variant subtyping (\cite{sangiorgi.walker:the-pi-calculus}) one gets
a sound and complete encoding.


\section{Conclusions}\label{sec:concl}
We have defined a translation from a system of session types for the
$\pi$-calculus into Igarashi and Kobayashi's generic type system
(GTS). We have proved correspondence results between process
reductions in the two systems, and between typing derivations; we can
also apply the generic type soundness theorem. 

Therefore, the translations clarify the relationship between session
types and GTS, and provide an interesting application of GTS, which
can thus be used to support analysis techniques for sessions.

\medskip
A preliminary version of this work was presented at PLACES'08 (but was
not published). The translations presented in that work were more
complicated (using forwarders) and did not consider subtyping.

\paragraph{Achievements.}
The encodings are a contribution: although encodings in process
calculi have been thoroughly studied, the two aspects presented herein
are novel: we are not aware of other investigations on how to encode
polarities or labels (as constants) in calculi without such
constructs. Our proposals may be used in other contexts.
%

The translation of branching and select types in the presence of
subtyping reveals a difficulty: the protocol of sending the labels of
the branch type to then offer a choice does not respect
the generic process subtyping discipline. This is because
 the encoding of labelled choice requires the labels to be passed as fresh
names; however, as subtyping does not allow changing the length of the
sequence of names passed, session types related by subtyping are
translated into unrelated process types. Even if the labels are passed
one by one (as in the encoding of polyadic into monadic
$\pi$-calculus), the encoded session types would not be related by
subtyping; they may, nonetheless, be related by simulation.

We do not see how to achieve a correspondence between subtyping in
session types and process types, but we leave the possibility of
proving a negative result for future work.

\medskip
An encoding using records (or a case constructor) and variant types in
the target language is, however, simple to achieve, as done by Dardha
\ea~\cite{DardhaEtal:st-revisited} and by Demangeon and
Honda~\cite{DemangeonHonda:fullyabstractsubtypedlinearpi}.

\paragraph{Assessment.}
In our opinion, the translation into GTS stresses that session types
themselves remain of great interest for programming language
design. Dardha \ea reached similar conclusions.
There are several reasons for that.

First, session types are a high-level abstraction for structuring
inter-process communication \cite{TakeuchiK:intblt}; preservation of
this abstraction and the corresponding programming primitives is very
important for high-level programming. The translation of a session
type into a process type produces a less informative type: the global
specification of behaviour is lost.

Second, there is now a great deal of interest in session types for
languages other than the $\pi$-calculus. Applying GTS would require
either translation into $\pi$-calculus, obscuring distinctive
programming abstractions, or the extension of GTS to other languages,
which might not be easy (apart from adding the constructors required
for subtyping).

Third, proofs of type soundness for session types are conceptually
fairly straightforward, even when these are liveness properties, as is
frequently the case. The amount of work saved by using the generic
type soundness theorem is relatively small. It may, however, be
possible to use treatments of deadlock-freedom in the generic type
system as a basis for understanding how to combine session types and
deadlock-freedom more directly.

Fourth, for practical languages we are very interested in typechecking
algorithms for session types; GTS does not yield an algorithm
automatically, so specific algorithms for session types need to be
developed in any case.

Fifth, the subtyping principles of session types, which provide
flexibility both for programming and for typing, are not easily
captured in a subtyping relation on ``plain'' process types. 

\paragraph{In short,} session constructors and session types are
encodable in process types, and one may use the power of GTS to
represent and reason about session specifications. As future work, we
plan to investigate concrete analysis techniques for sessions based on GTS.


\myparagraph{Acknowledgements.}
We are grateful to Lu\'{\i}s Caires, Kohei Honda, and Naoki Kobayashi
for useful discussions. We are also grateful for the careful
work and important comments/suggestions of various anonymous
reviewers.

Ant\'{o}nio Ravara is partially supported by the
Portuguese Fun\-da\-\c{c}\~{a}o para a Ci\^{e}ncia e a Tecnologia
via project ``Centro de Inform\'atica e Tecnologias da Informa\c{c}\~{a}o
(CITI/FCT/UNL)'' --- grant PEst-OE/EEI/UI0527/2014, and project
``Liveness, statically'' --- grant PTDC/EIA-CCO/117513/2010.
Simon Gay is partially supported by EPSRC grant EP/K034413/1 (From
Data Types to Session Types: A Basis for Concurrency and
Distribution).
%

%


\bibliographystyle{eptcs}
\bibliography{main}

\appendix
\begin{myfigure}
\centering\setlength\lineskip{10pt}
\AxiomC{$\repl\pinull \equiv \pinull$\ \ (\scshape SP-Star)}
\DisplayProof\hfil
\AxiomC{$(P | \pinull) \equiv P$\ \ (\scshape SP-Nil)}
\DisplayProof\hfil
\AxiomC{$\repl P \preceq (\repl P | P)$\ \ (\scshape SP-Rep)}
\DisplayProof\hfil\\[.5em]
\AxiomC{$(P | Q) \equiv (Q | P)$\ \ (\scshape SP-Commut)}
\DisplayProof\hfil
\AxiomC{$(P | (Q | R)) \equiv ((P | Q) | R)$\ \ (\scshape SP-Assoc)}
\DisplayProof\hfil\\[.5em]
\AxiomC{$(\res{\tilde x}P | Q) \equiv \res{\tilde x}(P | Q)$\ if $\tilde x$
  are not free in $Q$\ (\scshape SP-New)}
\DisplayProof\hfil\\
\AxiomC{$P \preceq P'$}
\AxiomC{$Q \preceq Q'$}
\rrulename{SP-Par}
\BinaryInfC{$(P | Q) \preceq (P' | Q')$}
\DisplayProof\hfil
\AxiomC{$P \preceq Q$}
\rrulename{SP-CNew}
\UnaryInfC{$\res{\tilde x}P \preceq \res{\tilde x}Q$}
\DisplayProof\hfil
\caption{Structural pre-order --- rules}
\label{fig:struct-pre}
\end{myfigure}


\section{Properties of the source and target languages}
\label{app:results-languages}

\paragraph{On session processes.}

\begin{definition}[from \cite{GaySJ:substp}]\label{def:balG}
  A type environment is \emph{balanced} if $\set{x^+, x^-}
  \subseteq \dom\Delta$ implies $\Delta(x^+)=\overline{\Delta(x^-)}$.
\end{definition}

\begin{theorem}[Type Preservation, from \cite{GaySJ:substp}]\label{thm:tp}
If $\Delta\vdash P$ with $\Delta$ balanced and $P\rightarrow P'$ then
there exists a balanced $\Delta'$ such that $\Delta'\vdash P'$.
\end{theorem}

Type safety guarantees that a correctly-typed process
contains no immediate possibilities for an error (in this
context the property is called ``session fidelity''). With
Type Preservation, it ensures that ``well-typed processes do not go wrong''.

\begin{theorem}[Session Fidelity, from \cite{GaySJ:substp}]\label{thm:sf}\label{thm:sf-full}
  Let $\Delta\vdash P$ for a balanced $\Delta$.
  \begin{itemize}
  \item If $P \equiv \res{\tilde x}(x^p\rcv{z}.P_1 \pars x^{\overline
      p}\send{y^q}.P_2 \pars Q)$ then $\set{x^p:\rcv{T}.S,x^{\overline
        p}\send{T}.\dual S,y^q:T} \subseteq \Delta,\tilde x:\tilde T$,
    and moreover, if $p \in \set{-,+}$ then $\set{x^+, x^-} \cap \fn Q
    = \0$.
  \item If $P \equiv \res{\tilde x}(x^p\branch{l_i:P_i}_{i \in
      I} \pars x^{\dual p}\sel l.Q \pars R)$ then $p\in\set{-,+}$, $l
    \in I$ and $\set{x^+, x^-} \cap \fn R = \0$.
  \end{itemize}
\end{theorem}

\paragraph{On generic processes.}

\begin{definition}[Correspondent properties]
  An invariant predicate $p$ on processes \emph{correspond to} a
  consistency predicate \ok\ on process types, if whenever $\okg$ and
  $\Gamma\triangleright P$ then $p(P)$ holds.
\end{definition}

\begin{theorem}[Adapted from Theorem 4.1.2 of
  \cite{IgarashiA:gentspjournal}]\label{thm:ts}
  Let some invariant $p$ on processes correspond to an \ok\ predicate.
  If $\Gamma\triangleright P$ and $\okg$, then $p(Q)$ holds for every
  $Q$ such that $P \to^* Q$.
\end{theorem}


\section{Subtyping}\label{app:subtyping}
\paragraph{Subtyping on session types.} Recall the subtyping relation
on session types (in Figure~\ref{fig:sess-sub}). Gay and Hole proved
that subtyping is a preorder.

To incorporate subtyping into the source language, we modify the rules
\textsc{(T-In)}, \textsc{(T-Out)}, and \textsc{(T-Offer)}. The new
versions of these rules are in Figure~\ref{fig:sess-rules}.
\begin{myfigure}
\centering\setlength\lineskip{5pt}
\AxiomC{}
\rrulename{S-End}
\UnaryInfC{$\typeend\subsess\typeend$}
\DisplayProof\hfil
\AxiomC{$S_1\subsess S'_1$}\AxiomC{$S_2\subsess S'_2$}
\rrulename{S-In}
\BinaryInfC{$\rcv{S_1}.S_2\subsess\rcv{S'_1}.S'_2$}
\DisplayProof\hfil
\AxiomC{$S'_1\subsess S_1$}\AxiomC{$S_2\subsess S_2'$}
\rrulename{S-Out}
\BinaryInfC{$\send{S_1}.S_2\subsess\send{S'_1}.S'_2$}
\DisplayProof\hfil
\AxiomC{$I\subseteq J$}\AxiomC{$\forall i\in I, S_i\subsess S'_i$}
\rrulename{S-Branch}
\BinaryInfC{$\echoice{l_i : S_i}_{i\in I}\subsess\echoice{l_i : S'_i}_{i\in J}$}
\DisplayProof\hfil
\AxiomC{$J\subseteq I$}\AxiomC{$\forall i\in J, S_i\subsess S'_i$}
\rrulename{S-Choice}
\BinaryInfC{$\ichoice{l_i : S_i}_{i\in I}\subsess\ichoice{l_i : S'_i}_{i\in J}$}
\DisplayProof\hfil
\caption{Subtyping for session types}
\label{fig:sess-sub}
\end{myfigure}


\begin{myfigure}
\centering\setlength\lineskip{5pt}
\AxiomC{$\Delta, x^p : S_1, y : S_2'\vdash P$}
\AxiomC{$S_2\subsess S'_2$}
\rrulename{T-In}
\BinaryInfC{$\Delta, x^p : \rcv{S_2}.S_1 \vdash x^p\rcv y.P$}
\DisplayProof\hfil
\AxiomC{$\Delta, x^p : S_1\vdash P$}
\AxiomC{$S'_2\subsess S_2$}
\rrulename{T-Out}
\BinaryInfC{$(\Delta, x^p : \send{S_2}.S_1) + y^q : S'_2 \vdash x^p\send y^q.P$}
\DisplayProof\hfil
\AxiomC{$I\subseteq J$}
\AxiomC{$\forall i\in I, (\Delta, x^p : S_i\vdash P_i)$}
\rrulename{T-Offer}
\BinaryInfC{$\Delta,x^p:\echoice{l_i : S_i}_{i\in I}\vdash x^p\branch{l_i:P_i}_{i\in J}$}
\DisplayProof\hfil
\caption{Session typing rules with subtyping}
\label{fig:sess-rules}
\end{myfigure}


\paragraph{Subtyping on process types.} We extend the subtyping
relation of the target language to take into account subtyping in
input and output process types. The new rules are in
Figure~\ref{fig:gen-sub-ext}. It is simple to check that the relation
is still a preorder.
\begin{myfigure}
\centering\setlength\lineskip{5pt}
{\scshape
%
\AxiomC{$\Gamma\subgen\Gamma'$}
\rulename{Sub-Out}
\UnaryInfC{$x\send\tau.\Gamma\subgen x\send\tau.\Gamma'$}
\DisplayProof\hfil
\AxiomC{$\Gamma\subgen\Gamma'$}
\rulename{Sub-In}
\UnaryInfC{$x\rcv\tau.\Gamma\subgen x\rcv\tau.\Gamma'$}
\DisplayProof\hfil
}
\caption{Generic subtyping relation: additional rules for input and output}
\label{fig:gen-sub-ext}
\end{myfigure}



\end{document}